\documentclass[iop]{emulateapj}
\usepackage{amsmath}
\usepackage{amssymb}
\usepackage{mathbbol}
\usepackage{dsfont}

\DeclareGraphicsExtensions{.eps,.eps.gz,.epsi}

\newcommand{\teff}{\mbox{$T_{\rm eff}$}} \newcommand{\logg}{{\rm{log}~$g$}}
\newcommand{\feh}{{\rm [Fe/H]}} 
\newcommand{\ebv}{$E(B-V)$}

\shorttitle{Stellar loci III: Photometric metallicities for half million FGK stars of Stripe 82}
\shortauthors{Yuan et al.}
\begin{document}
%\begin{CJK*}{UTF8}{gbsn}

\title{Stellar loci III: Photometric metallicities for half million FGK stars of Stripe 82}

\author
{Haibo Yuan\altaffilmark{1, 2},
Xiaowei Liu\altaffilmark{3, 1},
Maosheng Xiang\altaffilmark{3},
Yang Huang\altaffilmark{3},
Bingqiu Chen\altaffilmark{3, 2}
}
\altaffiltext{1}{Kavli Institute for Astronomy and Astrophysics, Peking University, Beijing 100871, P. R. China; email: yuanhb4861@pku.edu.cn, x.liu@pku.edu.cn}
\altaffiltext{2}{LAMOST Fellow}
\altaffiltext{3}{Department of Astronomy, Peking University, Beijing 100871, P. R. China}
%\end{CJK*}

\journalinfo{submitted to ApJ}
\submitted{Received ; accepted }

\begin{abstract}
We develop a method to estimate photometric metallicities
by simultaneously fitting the dereddened colors $u-g$, $g-r$, $r-i$ and $i-z$ from the SDSS 
with those predicted by the metallicity-dependent stellar loci.
The method is tested with a spectroscopic sample of main-sequence stars in Stripe 82 
selected from the SDSS DR9 and three open clusters.
With 1 per cent photometry, the method is capable of delivering photometric metallicities precise to about 0.05, 0.12, and 0.18\,dex
at metallicities of 0.0, $-$1.0, and $-2.0$, respectively,
comparable to the precision achievable with low-resolution spectroscopy at a signal-to-noise ratio of 10.
We apply this method to the re-calibrated Stripe 82 catalog and derive metallicities 
for about 0.5 million stars of colors $0.3 < g-i < 1.6$\,mag and distances between 0.3 -- 18 kpc.
Potential systematics in the metallicities thus derived, due to the contamination of giants and binaries, are investigated. 
Photometric distances are also calculated.
About 91, 72, and 53 per cent of the sample stars are brighter than $r$ = 20.5, 19.5, and 18.5\,mag, respectively.
The median metallicity errors are around 0.19, 0.16, 0.11, and 0.085\,dex for the whole sample,
and for stars brighter than $r$ = 20.5, 19.5, and 18.5\,mag, respectively. The median distance errors are 
8.8, 8.4, 7.7, and 7.3 per cent for the aforementioned four groups of stars, respectively. The data are publicly available. 
Potential applications of the data in studies of the distribution, (sub)structure, and chemistry of the Galactic 
stellar populations, are briefly discussed. The results will be presented in future papers.
\end{abstract}

\section{Introduction} 
Metallicity is one of the most important stellar parameters.
Together with the initial mass and age, they fully determine the current properties of a (single) star, 
including the internal structure and atmospheric spectrum.
The atmospheric metallicities of long-lived late-type stars 
retain a fossil record of the chemical composition of the interstellar medium at the time and place of their formation.
Measuring the metallicity distribution of stars in the Milky Way 
provides important clues to the formation, and chemical and dynamical evolution of the Galaxy 
(e.g., Casagrande et al. 2011; Schlesinger et al. 2012; An et al. 2013; Peng et al. 2013).

Metallicities, along with other stellar parameters, have now been measured 
for millions of stars from low- to medium- resolution spectra collected by multi-fiber spectroscopic surveys, 
including the Radial Velocity Experiment (RAVE; Steinmetz et al. 2006), 
the Sloan Digital Sky Survey (SDSS; York et al. 2000) and the LAMOST Galactic surveys 
(Deng et al. 2012; Zhao et al. 2012; Liu et al. 2014). 
With state-of-the-art stellar parameter pipelines such as the SEGUE Stellar Parameter Pipeline (SSPP; Lee et al. 2008a,b) developed for
the {\rm Sloan Extension for Galactic Understanding and Exploration} ({\rm SEGUE}; Yanny et al. 2009), 
the LAMOST Stellar Parameter Pipeline (LASP; Wu et al. 2014, in preparation), 
and the LAMOST Stellar Parameter Pipeline at Peking University (LSP3; Xiang et al. 2015), 
spectroscopic metallicities can be determined to an accuracy of about 0.1\,dex 
at good ($\gtrsim 10$) signal-to-noise ratios (SNRs). 
In spite of the high accuracy, of the hundreds of billions of Galactic stars spreading over the whole sky of 4$\pi$ steradians, 
the number that can be spectroscopically targeted within the limiting magnitudes of the surveys,
albeit now rapidly increased to order of a few millions thanks to the operation of LAMOST (Cui et al. 2012), remains to be small.
For example, the SDSS has hitherto spectroscopically surveyed hundreds of thousands of stars
selected with a variety of algorithms in several hundred pencil beams.
With 4,000 fibers, LAMOST has the potential to build a statically complete sample of millions of stars selected with 
a simple yet non-trivial algorithm (Yuan et al. 2015a), allowing the recovery of the underlying population for whatever class of objects
that are revealed by the spectra (e.g., Rebassa-Mansergas et al. 2014). 
Even so, various selection effects have still to be properly taken into account, which is a non-trivial task.
 
Stellar colors depend mainly on star's effective temperature.
However,  they also vary significantly with metallicity, in particular those blue colors as most 
metal absorption lines are in the blue wavelength range.
Therefore, photometric colors have long been used to estimate stellar metallicities, for example, 
in the traditional UV excess method (Wallerstein 1962; Sandage \& Smith 1963). 
For FG main-sequence (MS) stars of the same $B-V$ color, those of bluer $U-B$ colors have lower metallicities. 
The method has been extended to the SDSS photometric system by Ivezi{\'c} et al. (2008) using the SDSS DR6 data.
In spite of successfully mapping the metallicity distribution for a complete volume-limited sample of 
2 million FG stars at distances between 500 pc and 8 kpc,
the photometric metallicity estimators developed by Ivezi\'{c} et al. have some limitation and room of refinement: 
a) The estimators are only applicable to FG stars of $g-r$ colors between 0.2 -- 0.6\,mag; 
b) The calibration of \feh~estimates yielded by the SSPP has improved significantly since DR7, particularly for those very metal-rich 
and very metal-poor stars (Lee et al. 2008a, 2008b; Allende Prieto et al. 2008; Smolinski et al. 2011);
c) The estimators are built on colors $u-g$ and $g-r$ only. 
Employing as many colors as possible can potentially tighten the constraints and reduce the errors of metallicity estimates.
For example, An et al. (2013) use the SDSS $ugriz$ photometry and empirically calibrated stellar isochrones to drive 
temperatures, metallicities, and distances of individual stars.

The repeatedly scanned equatorial Stripe 82  ($|{\rm Dec}| < 1.266\degr$, 20{\rm h}34{\rm m} 
$< {\rm RA} <$ 4{\rm h}00{\rm m}) has delivered accurate photometry 
for about one million stars in $u,g,r,i,z$ bands (Ivezi{\'c} et al. 2007).
This is the largest single uniform data set publicly available with optical photometry internally consistent at 1 per cent level, 
providing ``a practical definition of the SDSS photometric system" (Ivezi{\'c} et al. 2007).
Using the spectroscopic information of about 24,000 stars in the Stripe available 
from SDSS  Data Release 9 (DR9; Ahn et al. 2012), 
Yuan et al. (2015b) have further re-calibrated the Stripe 82 photometric catalog 
with an innovative spectroscopy based stellar color regression (SCR) method, achieving an unprecedented internal calibration accuracy of 
about 0.005, 0.003, 0.002, and 0.002\,mag in colors $u-g$, $g-r$, $r-i$, and $i-z$, respectively. 
By combining the spectroscopic information and the re-calibrated photometry of Stripe 82, 
Yuan et al. (2015c, hereafter Paper\,I) have constructed a large, clean sample of MS stars with 
well determined metallicities and extremely accurate colors  
to derive metallicity-dependent stellar loci in the SDSS colors and investigate their intrinsic widths. 

Compared to previously derived stellar color loci (e.g., Covey et al. 2007; Davenport et al. 2014; Chen et al. 2014),
metallicity-dependent stellar loci, as discussed in Paper\,I, have 
a number of potential applications that have already led to some interesting results.
For example, in Paper\,II (Yuan et al. 2015d), we propose a Stellar Locus OuTlier (SLOT) method and 
provide a model-independent estimate of the binary fraction for field FGK stars. 
In this third paper of the series, 
we develop a method to obtain accurate photometric metallicities 
for stars of a wide range of color, and apply the method to the re-calibrated Stripe 82 catalog.

The paper is organized as follows.  In Section 2, we introduce the photometric data
and method. Various tests of the method are presented in Section 3. The results are presented in Section\,4.
The summary is given in Section\,5, along with a brief discussion of some 
potential applications of the newly deduced metallicities. 

\section{Data and Method}
\subsection{Data}
The re-calibrated Stripe 82 catalog contains 1,006,849 targets.
All the targets have $g$, $r$, and $i$ band magnitudes, 
but only 593,510 targets of them have $u$ band magnitudes, given that 
many red stars are too faint to be detected in $u$ band.
Also there are 157 targets without $z$ band magnitudes.
The stars are firstly dereddened using the dust reddening map of
Schlegel et al. (1998; SFD98 hereafter) and the empirical reddening coefficients of Yuan et al. (2015b),
derived using a star pair technique (Yuan, Liu \& Xiang 2013).
Given that color $u-g$ is most sensitive to metallicity,  
only stars that have a dereddened color\footnote{All colors and magnitudes quoted  
refer to the dereddened values unless specified otherwise.} $0.55 \le g-i \le 1.2$\,mag and  
all $u, g, r, i$, and $z$ magnitudes are included in the current analysis.
This yields a sample of 469,112 stars in total.

Panels (a) -- (e) of Fig\,1 plot the photometric errors as a function of the observed (undereddened) 
magnitudes for the sample stars. 
For $u$ band, the photon counting noises start to dominate the errors at 
$u \ga 19.0$\,mag. The errors are about 0.01 mag at $u=19.0$\,mag and 0.1 mag at $u=22.0$\,mag.
For $g$, $r$, and $i$ bands, the errors are dominated by the calibration uncertainties and essentially 
constant, at the level of 0.006, 0.005, and 0.005 mag for stars brighter than about $g=20.0$, 
$r=19.0$, and $i=19.0$\,mag, respectively.
At the detection limits ($g \sim 22.5$\,mag, $r\sim 21.5$\,mag, and $i\sim 21.0$\,mag), 
the errors increase to about 0.05\,mag.
For $z$ band, the photon counting noises dominate at $z \ga 18.0$\,mag, and the errors
are about 0.01, 0.02, 0.05, and 0.1 mag at $z$ = 18, 19, 20, and 21\,mag, respectively. 
Panels (f) and (g) show the sample distributions in the ($g-i$) -- ($u-g$) and ($g-i$) -- ($g-r$) planes, respectively.
The stellar loci of metallicities \feh~= $-$2.0 and 0.0 as derived in Paper\,I are also over-plotted. 
Most stars in the sample have an \feh~value between $-$2.0 and 0.0.
Panels (h) and (i) show the sample distributions in the  ($g-i$) -- $r$ and ($u-g$) -- $r$ color-magnitude diagrams. 
Clearly, the sample is a $u$-band limited one and contains more blue stars than red ones.

\begin{figure*}
\includegraphics[width=180mm]{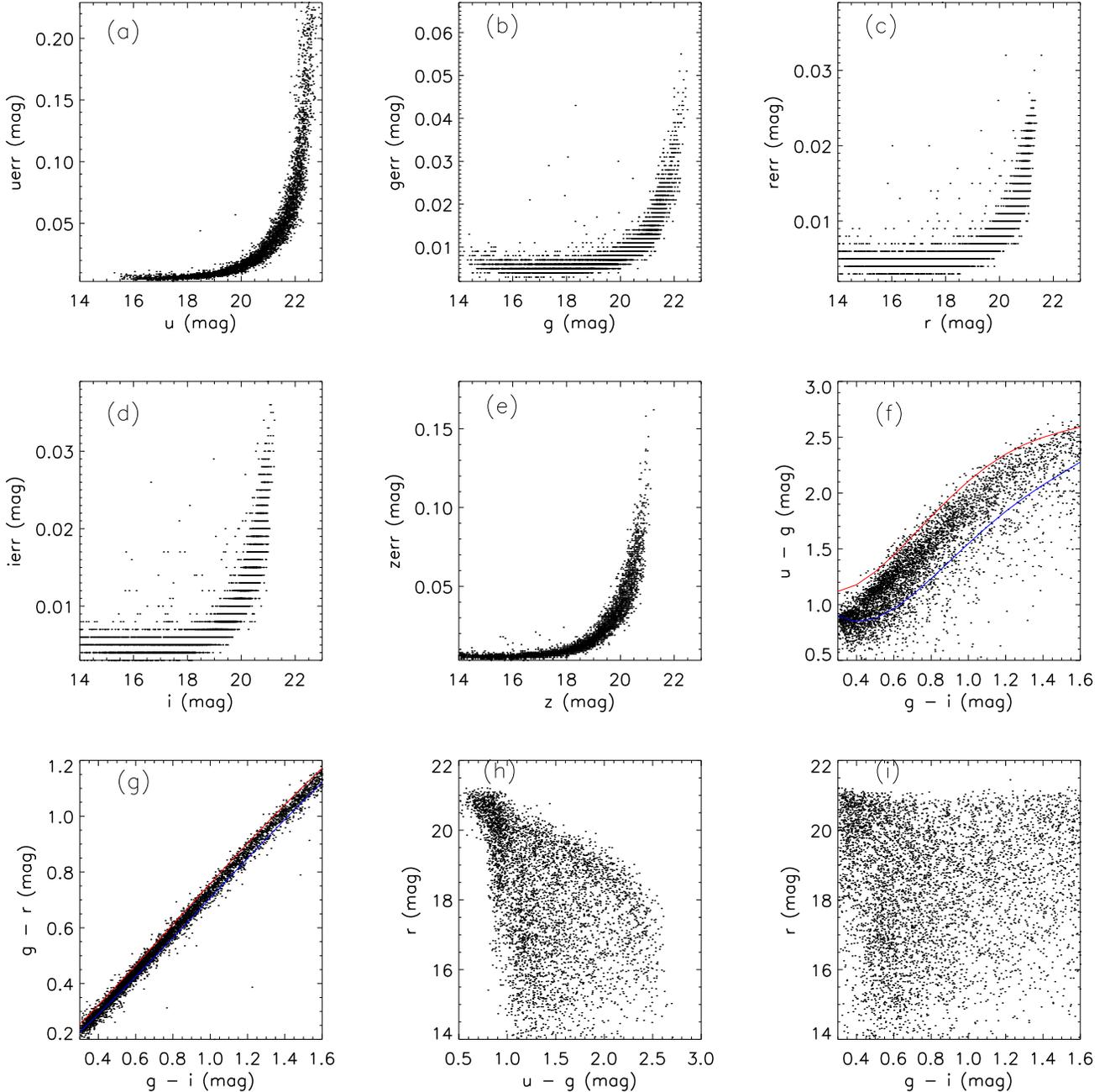}
\caption{
Distributions of photometric errors as a function of observed magnitudes [Panels (a) -- (e)] and  
distributions of sample stars in the ($g-i$) -- ($u-g$) (f), ($g-i$) -- ($g-r$) 
(g),  ($g-i$) -- $r$ (h), and ($u-g$) -- $r$ (i) planes.
In Panels (f) and (g), the blue and red lines denote the stellar loci for \feh~= $-$2.0 and 0.0, respectively.
To avoid crowdness, only one in a hundred randomly selected targets are plotted.}
\label{}
\end{figure*}

\subsection{Method}
We use a minimum  $\chi^2$ technique to determine metallicity.
Here $\chi^2$ is defined as:
\begin{align}
   \chi^2(\feh, g-i) = \nonumber\\  
\sum_{i=1}^{4}\frac{[c_{\rm obs}^i - R_{c}^{i} \times E(B-V) - c_{\rm int}^i(\feh, g-i) ]^2}{(\sigma_{c}^{i})^2 \times (4-2)}, 
\end{align}
where $c_{\rm obs}^i$ ($i$ = 1 -- 4) are the observed colors 
and $i = 1$, 2, 3, 4 for colors $u - g$, $g - r$, $r - i$ and $i - z$, respectively.
$R_{c}^{i}$, $c^i_{\rm int}$([Fe/H], $g-i$), and $\sigma^i_c$  (i = 1 -- 4) are respectively  
the reddening coefficients from Yuan et al. (2015b),
the intrinsic colors predicted by the metallicity-dependent stellar loci for a given set of metallicity \feh~and 
intrinsic color $g-i$, and the uncertainties of observed colors.
Values of \ebv~are from the SFD98 extinction map. 
$R_{c}^{i}$ = 1.060, 1.055, 0.604, and 0.528, for $i = 1$, 2, 3, and 4, respectively.  
Values of $\sigma_i$ are estimated from the magnitude errors and calibration uncertainties. 
The calibration uncertainties are adopted as 
0.005, 0.003, 0.002 and 0.002\,mag in colors $u-g$, $g-r$, $r-i$ and $i-z$, respectively (Yuan et al. 2015b).

We use a brute-force algorithm to determine the optimal \feh~and intrinsic $g-i$ color for each sample star.
For a given sample star, the value of \feh~is varied from $-2.5$ to 0.5 at a step of 0.01\,dex,
and the value of $g-i$ is varied from $(g-i)_{\rm obs}$ $-$ 2$\times$$\sigma_{g-i}$ to 
$(g-i)_{\rm obs}$ + 2$\times$$\sigma_{g-i}$ (mag) at a step
of 0.2$\times$$\sigma_{g-i}$\,mag. A 301$\times$21 array of $\chi^2$ values is 
calculated to find the minimum value of $\chi^2$,
$\chi^2_{\rm min}$, and the corresponding values of intrinsic $g-i$ color, $(g-i)_0$, and of metallicity \feh. 
The error of \feh~is estimated as $1\sigma$ uncertainty that corresponds to the difference of \feh~values at 
$\chi_{\rm min}^2$ and $\chi_{\rm min}^2$+1 (Avni 1976; Wall 1996).

\section{Tests of the method}

\subsection{Internal test}

We first apply the technique to the spectroscopic sample of Stripe 82 
used to construct the metallicity-dependent stellar loci in Paper\,I, 
and compare the resultant photometric metallicities with the spectroscopic values.
The spectroscopic sample of Paper\,I contains 24,492 stars of a spectral SNR $>$ 10, 
a line-of-sight extinction \ebv~$\le$ 0.15\,mag, a surface gravity \logg~$\ge$ 3.5\,dex, a color $0.3 \le g-i \le 1.6$\,mag,
an effective temperature \teff~$\ge$ 4,300\,K, and a metallicity $-2.0 \le \feh~\le 0.0$.
In the current work, we have relaxed the cut in \feh~ and included another 1,173 stars with 
a \feh~value either larger than 0.0 or lower than $-$2.0 to allow us test the validity range of \feh~of the method. 
The final test sample thus contains 25,665 stars.

The top left panel of Fig.\,2 shows a histogram distribution of $\chi_{\rm min}^2$ values of the sample.
Only 11.4, 4.8, and 1.3 per cent of the stars show a $\chi_{\rm min}^2$ value larger than 3, 5, and 10, 
respectively. Most stars showing a large $\chi_{\rm min}^2$ value are probably binaries, as 
binaries generally do not follow the stellar loci of single (MS) stars (Yuan et al. 2015d).
The top right panel of Fig.\,2 displays a comparison of the 
photometric metallicities thus derived and the corresponding spectroscopic values taken from the DR9. 
There is no significant systematic difference, and the dispersion is only 0.15\,dex.
The random errors of \feh~in the DR9 as yielded by the SSPP pipeline have 
been determined as a function of spectral SNR and \feh~for FGK MS stars using duplicate observations (Yuan et al. 2015d).
The random errors of \feh~for the test sample stars range from 0.03 to 0.20\,dex, with 
a median value of 0.08\,dex. 
The result suggests that the photometric metallicities derived in the current work have a typical error of 0.13\,dex.
Note that for SNR = 10, the random errors of \feh~yielded by the SSPP 
are about 0.11, 0.16, and 0.19\,dex at \feh~= 0, $-$1, and $-$2, respectively.
The bottom two panels of Fig.\,2 display the differences between the photometric and spectroscopic metallicities
as a function of metallicity \feh~and color $g-i$.
The sample is divided into bins of \feh~and $g-i$. The median and dispersion values 
for individual bins are over-plotted. 
No systematic dependences of the differences on \feh~are found within the \feh~range 
of $-2.0$ -- 0.0, and on color $g - i$ within the range of the plot.
Not surprisingly, at \feh~$>$ 0.0, the photometric metallicities are under-estimated by about 0.15\,dex.
At \feh~$<$ $-2.0$, the photometric metallicities are over-estimated by about 0.2\,dex.
Note the boundary in the top left corner of the bottom left panel corresponds to the lower limit 
of photometric metallicities treated in the current work, i.e., $-2.5$.
For stars of $g-i$ $\lesssim$ 0.35\,mag and \feh~ $\lesssim$ $-$1.5,  
the metallicity-dependent stellar loci of different metallicities intersect, leading to 
approximately constant \feh~value around $-2.2$.
This artifact can be clearly seen in the bottom left panel of Fig.\,2.
This is also the reason why the bottom right panel shows large scatters at the bluest colors.

The uncertainties of photometric metallicity depend on the apparent magnitude, metallicity and color of the star under consideration.
For the test sample, the median uncertainty is 0.10\,dex and the mean value is 0.13\,dex. 
This is consistent with the typical error of 0.13\,dex as yielded by 
comparing the photometric and spectroscopic metallicities.
Fig.\,3 plots the estimated uncertainties of photometric \feh~of the 
test sample against the uncertainties of color $u-g$, against the spectroscopic 
\feh~from the DR9, and color $g-i$, respectively. 
A good, linear correlation is seen between the uncertainties of
photometric \feh~and the errors of $u-g$, as can be expressed as, 
\begin{equation} %\[
   \sigma({\rm [Fe/H]}) = 4.25\times \sigma (u-g) + 0.034.
\end{equation}
A 0.02\,mag error in $u-g$ color yields a 0.12\,dex uncertainty in \feh, 
in line with the variation of $u-g$ color to the metallicity (Paper\,I).
The \feh~uncertainty is anti-correlated with \feh,
\begin{equation} %\[
   \sigma ({\rm [Fe/H]}) = -0.074 \times {\rm [Fe/H]} + 0.043.
\end{equation}
Such an anti-correlation is expected.
For a given uncertainty of $u-g$, the stellar loci are more sensitive to metallicity at higher metallicities (Paper\,I).
Note that the constants in Eqs.\,(2) and (3) are valid only for the test sample, for which the median error in $u-g$ color is 
about 0.013\,mag and the median \feh~is $-0.71$. 
The right panel of Fig.\,3 illustrates that photometric estimation of metallicity works best 
for stars of $g-i$ colors between 0.4 -- 1.2\,mag. 
Note also that for individual stars, the estimated errors in some cases could be significantly under-estimated. 

\begin{figure}
\includegraphics[width=90mm]{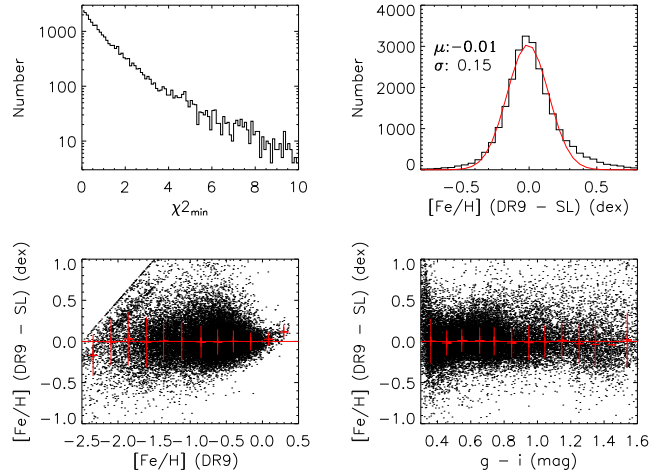}
\caption{
{\em Top panels:} Histogram distributions of $\chi_{\rm min}^2$ values of the test sample (left)
and the differences of photometric metallicities derived in the current work 
and the corresponding spectroscopic values from the DR9 (right). 
The red curve denotes a Gaussian fit, with the mean and dispersion values labeled.
{\em Bottom panels}: Differences between the photometric and spectroscopic metallicities are 
plotted as a function of spectroscopic metallicity (left) and of $g-i$ color (right). 
The sample is divided into bins of \feh~and $g-i$ of widths 0.25\,dex and 0.1\,mag, respectively. 
For each bin, the median and dispersion values are over-plotted in red.} 
\label{}
\end{figure}

\begin{figure}
\includegraphics[width=90mm]{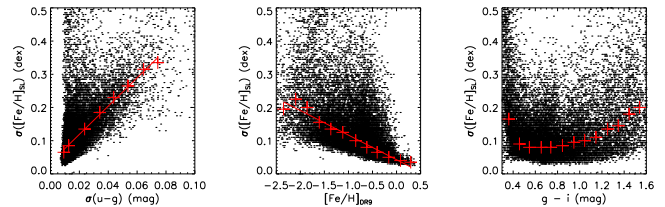}
\caption{
Uncertainties of photometric \feh~of the test sample plotted against errors of $u-g$ color (left), 
spectroscopic values of \feh~from DR9 (middle), and color $g-i$ (right). 
The sample is divided into bins of $\sigma (u-g)$, \feh, and $g-i$ of widths 0.01\,mag, 0.25\,dex, and 0.1\,mag, 
respectively. The median and dispersion values of the photometric estimates of [Fe/H] of the individual bins 
are over-plotted in red. The red lines in the left and middle panels denote linear fits to those median values.
}
\label{}
\end{figure}

\begin{figure}
\includegraphics[width=90mm]{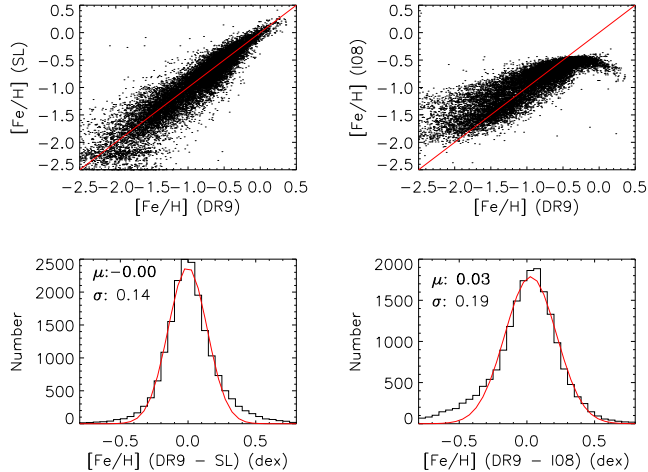}
\caption{
Comparisons of photometric metallicities derived in the current work (left panels) 
and those using the relation of Ivezi{\'c} et al. (2008) (right panels) with the spectroscopic values from the DR9
for a sub-sample of test stars of $0.2 < g-r < 0.6$\,mag.
The diagonal lines in the top panels denote equality of the two quantities in comparison.
The red curves in the bottom panels denote Gaussian fits to the distributions of the differences, 
with the mean and dispersion values labeled.
}
\label{}
\end{figure}

For stars of the test sample of $g-r$ color between 0.2 -- 0.6\,mag, we have also determined their  
photometric metallicities using Eq.\,(4) of Ivezi{\'c} et al. (2008).
Fig.\,4 compares the results from the current work, those deduced using the relation 
of Ivezi{\'c} et al., and the spectroscopic values from the DR9 for this sub-sample of stars.
Compared to the relation of Ivezi{\'c} et al., out current method not only yields much smaller 
discrepancies (0.14 versus 0.19 dex on average) when comparing with the spectroscopic values, 
but also is applicable to a much wider color range. The strength of the current method 
comes from the fact that it makes use of as much information as available. 
The systematics of results yielded by the relation of Ivezi{\'c} et al. at metallicities 
higher than $-$0.5 or lower than $-$1.5 is possibly 
caused by the potential metallicity calibration problems in the SDSS DR6. 

\subsection{Test with star clusters}
We have applied the method to three open clusters (OCs; NGC\,2420, M\,67, and NGC\,6791),
using the SDSS photometry from An et al. (2008).
The \feh~estimates of the OCs NGC\,2420, M\,67, and NGC\,6791 
have been previously determined by spectroscopy at $-$0.37 (Anthony-Twarog et al. 2006),
0.0 (An et al. 2007), and 0.47 (Carretta et al. 2007), respectively.
The candidate member stars are selected based on the spatial positions as well as 
the distributions in the $g-i$ versus $r$ color-magnitude diagram (Xiang et al. 2015).
The stars are then dereddened using the SFD extinction map and the empirical reddening coefficients of Yuan, Liu \& Xiang (2013).
Only MS stars are selected.

The results are showed in Fig\,5.
For NGC\,2420, the mean photometric metallicity is $-0.29\pm0.13$, 0.02\,dex higher than the mean spectroscopic
value from the DR9 and 0.08\,dex larger than the literature value.
The photometric metallicities show some trend with $g-i$ color in the sense that the values
are somewhat higher for stars bluer than $g-i < 0.6$\,mag. At $g-i > 0.6$\,mag, the photometric estimates agree
very well with the literature. The photometric metallicities follow the spectroscopic values from the DR9,
suggesting that the over-estimates at blue colors are possibly caused by calibration problems of the DR9 for blue, metal-rich stars.
M\,67 shows a similar trend. Its mean photometric metallicity is $-0.16\pm0.11$,
0.07\,dex lower than the mean spectroscopic value from the DR9 and 0.16\,dex lower than the literate value.
For NGC\,6791, the mean photometric metallicity is $0.28\pm0.16$, 0.08\,dex higher than
the mean spectroscopic value from the DR9 and 0.19\,dex lower than the literature value.
The small scatters of the photometric estimates of metallicities for the three OCs as well as the small systematic 
discrepancies with the literature values suggest that the photometric method of the current work is capable of 
providing metallicity estimates for MS stars accurate to 0.1 -- 0.15\,dex.

\begin{figure*}
\includegraphics[width=180mm]{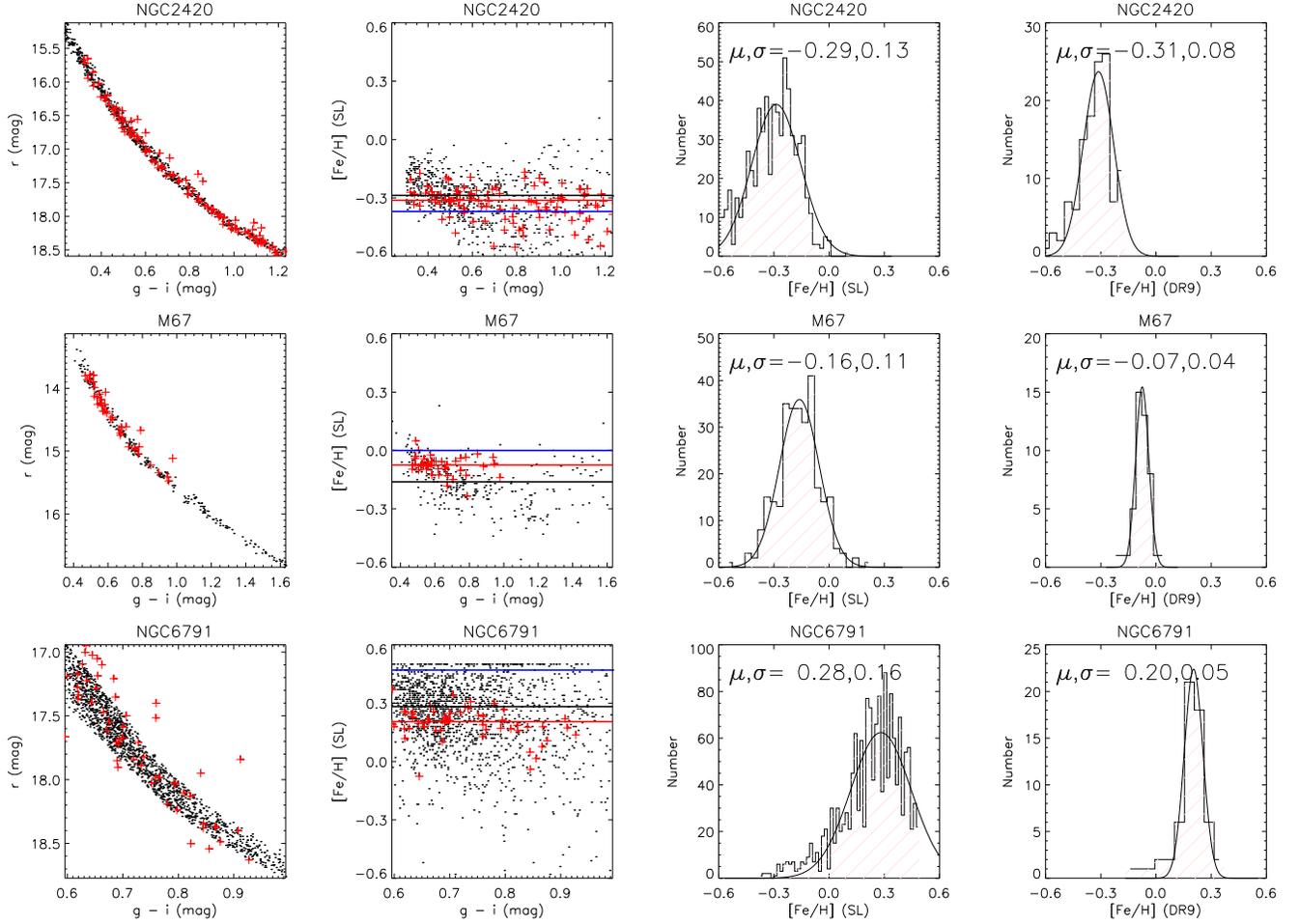}
\caption{
Comparison of photometric metallicities of MS stars of OCs NGC\,2420 (upper panels), M\,67 (middle panels), and NGC\,6791 (lower panels) with other estimates.
{\em Column 1:}\, ($g-i$) versus $r$ color-magnitude diagram of the selected MS stars in the three OCs.
Red pluses denote member stars spectroscopically targeted by the SDSS.
Note the outliers in the  ($g-i$) versus $r$ diagram of M\,67 are binaries.
{\em Column 2:}\, Photometric metallicities plotted against color $g-i$.
Red pluses denote member stars spectroscopically observed by the SDSS.
The black, red, and blue lines
denote the mean photometric metallicities, the mean spectroscopic metallicities from the DR9,
and the literature values, respectively.
{\em Column 3:}\, Histogram distributions of photometric metallicities.
{\em Column 4:}\, Histogram distributions of spectroscopic metallicities from SDSS DR9.
In the 3rd and 4th columns, also over-plotted are Gaussian fits to the distributions,
with the central values and dispersions of the fits labeled.
}
\label{}
\end{figure*}

\subsection{Effects of giant stars}
The metallicities presented in the current work are estimated using metallicity-dependent stellar loci constructed from MS stars. 
To investigate the possible systematics due to the contamination of giants, we apply the method to 
a sample of 1,996 giant stars selected from the Stripe 82 that are spectroscopically targeted by the SDSS and  
have a line-of-sight extinction \ebv~$\le$ 0.15\,mag, \logg~$\le$ 3.5\,dex, $0.55 \le g-i \le 1.6$\,mag,
\teff~$\ge$ 4,300\,K, and $-2.5 \le \feh~\le 0.5$.
The resultant photometric metallicities are then compared to the spectroscopic values.

The results are shown in Fig.\,6.
Compared to results of the MS sample, 
the $\chi_{\rm min}^2$ values of the giant sample are generally larger. 
There are 33.2, 16.4, and 3.9 per cent giant stars of $\chi_{\rm min}^2$ values larger than 3, 5, and 10, respectively.  
The deduced photometric metallicities are systematically higher than the spectroscopic values by an average value of 0.13\,dex, with a 
large dispersion of 0.25\,dex, which is significantly higher than the value dispersion found for the MS sample.
The systematics show small variations with $g-i$ color, but depend strongly on \feh.
The systematic differences are close to zero for \feh~between $-1.0$ and $-0.5$, and increase to 0.5\,dex at \feh $= -2.0$.
The results suggest that the stellar loci of giant stars are different from those of MS stars, especially for metal-poor stars. 
In the fourth paper of this series (Yuan et al. 2015e), we will construct the metallicity dependent stellar loci for red giants, 
use the results to discriminate between red giants and MS stars, and to estimate photometric metallicities of red giants based on the SDSS photometry.

\begin{figure}
\includegraphics[width=90mm]{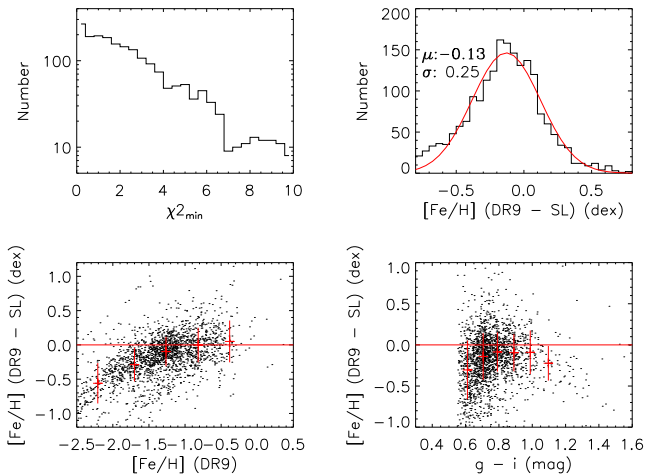}
\caption{
Same as to Fig.\,2 but for the sample of giant stars.
}
\label{}
\end{figure}

\subsection{Effects of binaries}
Binaries are ubiquitous in the Galaxy (e.g., Raghavan et al. 2010), 
and the binary fraction varies with spectral type and possibly also with metallicity (e.g., Yuan et al. 2015d). 
Because the colors of MS binaries generally deviate from the metallicity-dependent 
stellar loci of single MS stars (Yuan et al. 2015d), 
estimates of photometric metallicities of MS stars could be affected by binarity. 
In this subsection, we explore such effects.

Following Yuan et al. (2015d), we simulate the combined colors of binary systems composed of two MS stars
for all possible combinations of primary stars of $g-i$ colors between 0.3 -- 1.6\,mag
and secondary stars of $g-i$ colors between 0.3 -- 3.0\,mag, 
on the condition that the color of the secondary is no bluer than that of the primary.
We further assume that the primary and secondary stars have the same \feh.
For a binary system of given $g-i$ color set and metallicity \feh, 
the absolute $r$-band magnitudes, $M(r)$, of the two component stars are
computed using the photometric parallax relation of Ivezi{\'c} et al. [2008; Eq.\,(A7)].
The $u-g$, $g-r$, $r-i$, and $i-z$ colors of the two components 
are computed using the metallicity-dependent stellar loci determined
in Paper\,I for $g-i \leq 1.6$\,mag and using the stellar loci of Covey et al. (2007) for $g-i > 1.6$\,mag.
The absolute $r$-band magnitudes and $u-g$, $g-r$, $r-i$, and $i-z$ colors are then used to derive the absolute magnitudes in $u,g,i,$ and $z$ bands.
Given the absolute magnitudes in $u,g,r,i,$ and $z$ bands of the two components,
the magnitudes and colors of the binary system are calculated.
Then we use the combined colors to derive the photometric metallicity of the binary system, assuming that 
the errors of 0.025, 0.015, 0.015, 0.02 and 0.015\,mag in colors $u-g$, $g-r$, $r-i$, $i-z$, and $g-i$, respectively.
The differences between the true and derived values of \feh~ 
are then deduced and plotted in Fig.\,7 as a function of the $g - i$ colors of the primary and secondary stars, respectively.
Three sets of simulations are performed, for \feh~=~0.0, $-1.0$, and $-2.0$, respectively.

Fig.\,7 shows that:
1) The photometric metallicities deduced for the simulated binaries are typically more metal-poor than 
the true values, consistent with the fact that the combined colors are typically bluer 
in $u-g$ and $g-r$ and redder in $r-i$ and $i-z$ than those predicted for single stars.
The systematics are in fact the underlying cause responsible for the asymmetric distributions of the differences 
between the photometric and spectroscopic metallicities of the test MS sample (c.f., top right panel of Fig.\,2);
2) The systematics are however relatively small. The median differences amount to only $-$0.11, $-$0.15, and $-$0.07\,dex 
for \feh~=~0.0, $-1.0$, and $-2.0$, respectively.
The maximum differences are about 0.35\,dex for \feh~=~0.0 and 0.5\,dex for \feh~=~$-1.0$ and $-2.0$.
As expected, the differences are close to nil 
when the primary and secondary stars have close to unity or very small mass ratios. The differences reach
the maximum values at certain combinations of the colors of the component stars.
The results suggest that MS binaries can mimic the colors of more metal-poor (single) stars.
%In some cases, 0.5\,dex. 
This effect should be kept in mind when searching for very metal-poor stars from photometry alone.
Note that the artifacts at $g-i$ $\sim$ 1.6\,mag of the secondary are caused
by the different sets of stellar loci used below and above this color.
The artifacts at the top right corner of each panel are due to the fact 
that the combined $g-i$ colors of the simulated systems become redder than 1.6\,mag, and are thus beyond 
the color range that can be dealt with the method presented in the current work.
Due to the problems in $u-g$ color of the metallicity-dependent stellar loci at $g-i <$ 0.4\,mag,
the simulated differences in \feh~are probably unrealistic at $g-i <$ 0.4\,mag.

\begin{figure*}
\includegraphics[width=180mm]{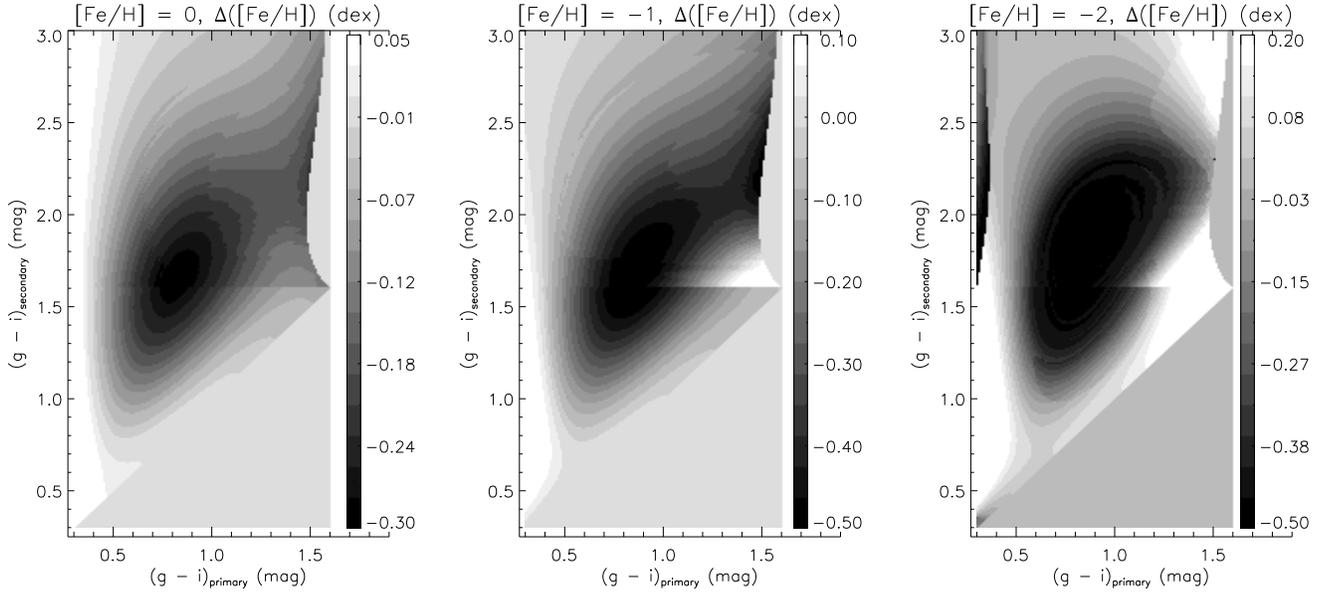}
\caption{
Distributions of the differences between the true and photometric estimates of metallicity
as a function of $g - i$ colors of the primary and secondary stars of the simulated binary systems. 
The binaries are assumed to compose of two MS single stars of the same metallicity of \feh~=~0.0 (left), $-1.0$ (middle), and $-2.0$ (right), respectively. 
A colorbar is over-plotted by the side in each case. The stellar loci of Paper\,I 
and those of Covey et al. (2007) are used for stars of $g-i \leq 1.6$ and $>$ 1.6\,mag, respectively.
The errors in colors $u-g$, $g-r$, $r-i$, $i-z$, and $g-i$ are assumed to be 0.025, 0.015, 0.015, 0.02, and 0.015\,mag, respectively.
}
\label{}
\end{figure*}

\section{Results} 
We now apply the technique to the whole selected photometric sample consisting of 
about 0.5 million stars of Stripe 82 and estimate their photometric metallicities. 
With the newly derived photometric metallicities,  
the photometric distances of the stars are then calculated using the photometric parallax relation of Ivezi{\'c} et al. [2008; Eq.\,(A7)].
The distance errors are also estimated, taking into account the uncertainties in $g-i$ colors, those in photometric metallicities,
and the intrinsic scatter of the photometric parallax relation (0.13 mag).
The data are accessible from http://162.105.156.249/site/Stripe82/, along with a descriptive readme.txt file.
About 91, 72, and 53 per cent of the sample stars are brighter than $r$ = 20.5, 19.5, and 18.5\,mag, respectively.
In the calculations, we have assumed that all stars are single dwarfs.
Using the Besan\c{c}con Galactic model (Robin et al. 2003), 
we estimate the fraction of giant stars (including sub-giants) in the sample
is about 4 per cent, in agreement with 
the estimate of Juri{\'c} et al. (2008). The fraction is slightly higher for brighter stars.
The binary fraction as a function of stellar colors and metallicities has been determined by Yuan et al. (2015d).
One should keep in mind the small contamination of giant stars and the possible effects of binaries when using the sample.
 
The top panel of Fig.\,8 plots histogram distribution of $\chi^2_{\rm min}$ for the sample.
For the whole sample, 20.0, 9.2, and 2.3 per cent of the stars show a $\chi^2_{\rm min}$ value larger than 3, 5, and 10, respectively.  
The fractions are about two times larger than the corresponding values of the test sample, due to the contamination of giant stars.
The fractions first decrease slightly towards brighter stars, with the corresponding percentage being 
18.0, 8.0, and 2.0 per cent for stars brighter than $r$ = 20.5\,mag
and being 16.2, 7.2, and 1.8 per cent for stars brighter than $r$ = 19.5\,mag.
However, the fractions then increase, probably due to a larger fraction 
of giants amongst bright stars, with the corresponding percentage being 
16.2, 7.5, and 2.1 per cent for stars brighter than $r$ = 18.5\,mag.

The top middle panel of Fig.\,8 shows that 
the metallicity distribution of the sample peaks around $-$0.45, and have an extended tail of metal-poor stars. 
Another peak around $-2.25$ is an artifact and caused by the systematics of the method for  
blue, metal-poor stars, as discussed in Section\,3.1.
The fraction of metal-poor stars decreases rapidly as the sample stars become brighter (closer).
The trend is also seen in Fig.\,9, which plots the photometric metallicities against $g-i$ colors for stars of different magnitude ranges.
Fig.\,9 also shows that stars of different colors exhibit different metallicity distributions, 
largely because stars of different colors have different intrinsic luminosities and thus probe different depths.
Note that the branch of stars in the bottom left corner is also an artifact and caused by the same systematics of the method.
The numbers of very metal-poor stars, defined as those of \feh~$< -2.0$ and $\chi^2_{\rm min}$ $<$ 3.0, 
are 210, 733, 2,487, and 9,636 to a limiting magnitude of $r$ = 16.5, 17.5, 18.5, and 19.5\,mag, respectively.
The top right panel of Fig.\,8 plots histogram distribution of the errors of photometric metallicities of the sample. 
The median values are 0.19, 0.16, 0.11, and 0.085\,dex for the whole sample, 
and for those stars that are brighter than $r$ = 20.5, 19.5, and 18.5\,mag, respectively.
About 37.6, 31.9, 16.7, and 3.4 per cent of the stars have metallicity errors larger than 0.3\,dex 
for the aforementioned four groups of stars, respectively.

The bottom two panels of Fig.\,8 plot histogram distributions of photometric distances and errors for the sample.
The maximum distance reached by the whole sample is about 18\,kpc, and decreases to 15, 10, and 6 kpc for 
stars brighter than $r$ = 20.5, 19.5, and 18.5\,mag, respectively.
The median distance errors are 8.8, 8.4, 7.8, and 7.3 per cent for the aforementioned four groups of stars, respectively.
For the four groups of stars, 
about 40.2, 35.9, 25.8, and 8.9 per cent of them have distance errors larger than 10 per cent, respectively,
and only 7.0, 5.8, 2.3, and 0.2 per cent of them have distance errors larger than 20 per cent, respectively.
Note that most distance errors of the sample stars are contributed by the intrinsic scatter of the photometric 
parallax relation of Ivezi{\'c} et al. (2008).
If one neglects the intrinsic scatter of the relation, the median distance errors reduce to 
6.1, 5.6, 4.6, and 3.8 per cent for the four groups of stars, respectively.
It implies that if one selects a sub-sample of stars of a narrow range in color and metallicity, 
then the errors of their (relative) distances can be quite small.
The distributions in the  $R$ -- $Z$ plane is shown in Fig.\,10.
The distributions are different for stars of different metallicity ranges, with the more metal-rich stars 
being closer to the Galactic plane. 

The sample also contains a number of targets of abnormal colors, exhibiting very large values of $\chi^2_{\rm min}$. 
Most of them are faint white-dwarf-main-sequence binaries. 
Giant stars and F turnoff stars of high metallicities, in spite of their small numbers in the sample, 
also show very large values of $\chi^2_{\rm min}$.

\begin{figure*}
\includegraphics[width=180mm]{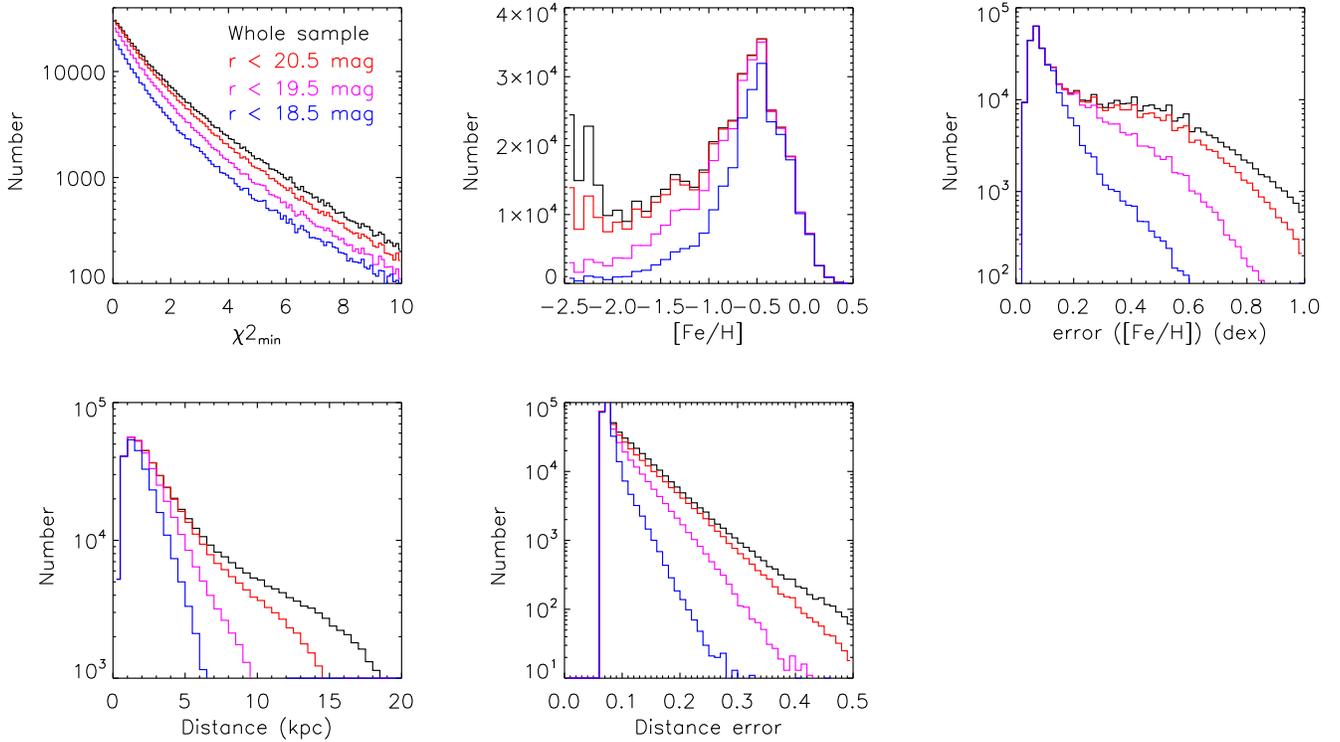}\caption{
Histogram distributions of $\chi^2_{\rm min}$ (top left), 
photometric metallicities (top middle) and errors (top right),
photometric distances (bottom left) and errors (bottom middle) 
for the whole photometric sample (black), and sample stars brighter 
than $r$ = 20.5 (red), 19.5 (purple), and 18.5\,mag (blue), respectively.  
}
\label{}
\end{figure*}

\begin{figure}
\includegraphics[width=90mm]{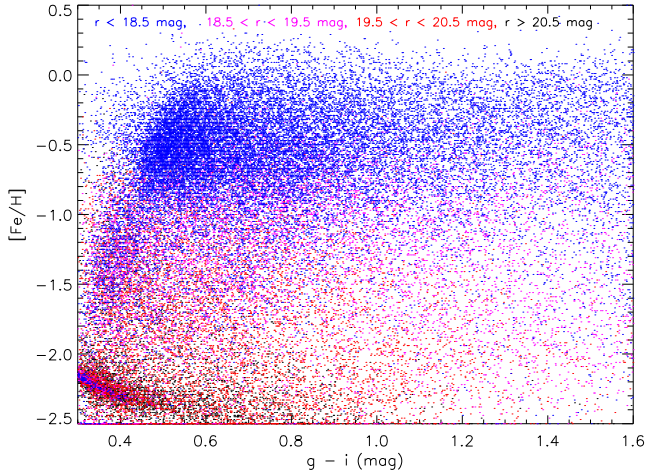}\caption{
Distribution in the ($g - i$) -- \feh~ plane of the photometric sample.
Different colors denote stars of different brightness, as marked near the top of the plot.
To avoid crowdness, only one-in-ten randomly selected targets are shown.
}
\label{}
\end{figure}

\begin{figure}
\includegraphics[width=90mm]{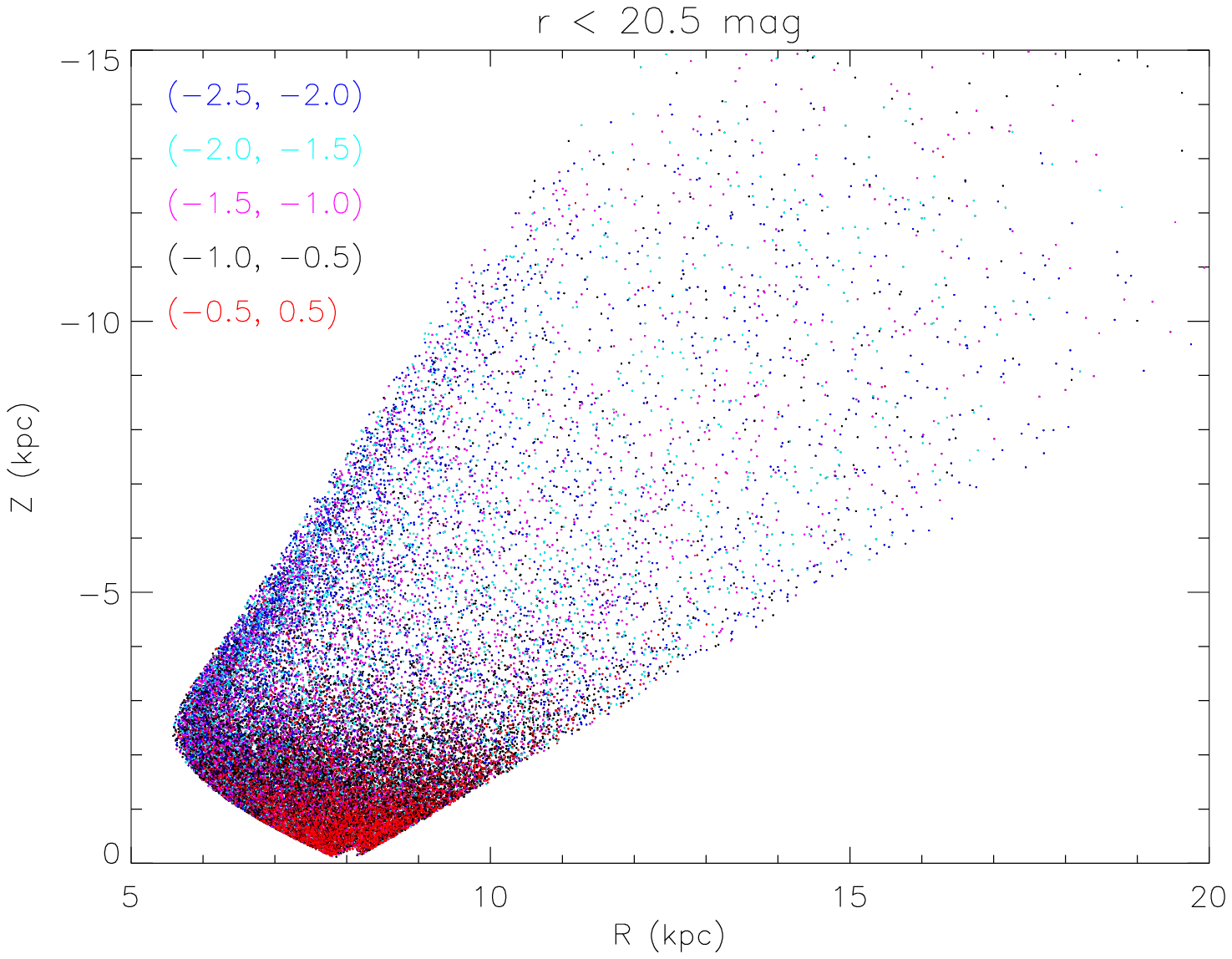}\caption{
Spatial distribution in the $R$ -- $Z$ plane of sample of stars brighter than $r$ = 20.5\,mag. 
The Sun is located at ($R$, $Z$) = (8.0, 0.0)\,kpc.
Different colors denote stars of different metallicity ranges, as marked in the bottom left corner of the plot. 
To avoid crowdness, only one-in-ten randomly selected targets are shown.
}
\label{}
\end{figure}

\section{Summary}
In this paper, we present a method to estimate photometric metallicities for MS stars, 
by simultaneously fitting the dereddened colors in $u-g$, $g-r$, $r-i$, and $i-z$ with 
those predicted by the metallicity-dependent stellar loci presented in Paper\,I.
The method is tested using a spectroscopic sample of MS stars of Stripe 82 selected from the SDSS DR9 and three 
open clusters imaged and spectroscopically targeted by the SDSS.
With 1 per cent photometry, the method is capable of delivering photometric metallicities precise to about 0.05, 0.12, and 0.18\,dex
at metallicities of 0.0, $-$1.0, and $-$2.0, respectively,
comparable to the precision achievable with low-resolution spectroscopy for a signal-to-noise ratio of 10.
Compared to the method of Ivezi{\'c} et al (2008), the current method is not only more accurate, 
but also applicable to a wider range of color and metallicity ($0.3 < g-i < 1.6$\,mag, $-2.5 <$ \feh~$<$ 0.5), 
thanks to the employment of all available colors and the improved metallicity calibration of the SDSS DR9.  
Systematics of the method due to the possible contamination of giant stars and MS binaries are also discussed. 
The photometric metallicities of giant stars, when derived using the stellar loci of MS stars, are systematically 
under-estimated at \feh~$<$ $-$1.0. The discrepancies become larger for stars of lower metallicities.
Similarly, the method under-estimates the metallicities of MS binaries, by a typical amount of 0.1\,dex, and up to 0.5\,dex in some extreme cases.
Those systematics should be kept in mind when using the sample to search for very metal-poor stars.

We apply the method to a photometric sample of about 0.5 million stars in the Stripe 82 region that have $u, g, r, i$, and $z$ 
magnitudes and $g-i$ colors between 0.3 -- 1.6\,mag.  
About 91, 72, and 53 per cent of the sample stars are brighter than $r$ = 20.5, 19.5, and 18.5\,mag, respectively.
The median metallicity errors obtained are 0.19, 0.16, 0.11, and 0.085\,dex for the whole sample, and for stars 
brighter than $R = 20.5$, 19.5 and 18.5\,mag, respectively.
The metallicity distribution of the sample peaks around $-$0.45, and has a tail extending to lower metallicities.
The numbers of very metal-poor stars, defined as those of \feh~$< -2.0$ and $\chi^2_{min}$ $<$ 3.0,
count 210, 733, 2,487, and 9,636 to a limiting magnitude of $r$ = 16.5, 17.5, 18.5, and 19.5\,mag, respectively.
With the newly deduced photometric metallicities, photometric distances are also calculated using 
the photometric parallax relation of Ivezi{\'c} et al. (2008).
The maximum distances probed by the aforementioned four group of stars are 
about 18, 15, 10, and 6 kpc, respectively. 
The median distance errors, estimated by taking into account contributions from the uncertainties of 
photometry metallicity, measurements of color $g-i$, and from the scatter of the photometric parallax relation, 
are 8.8, 8.4, 7.8, and 7.3 per cent for the four groups of stars, respectively.

With accurate photometric metallicities and distances, 
the sample in the current work provides an excellent, magnitude-limited data-set to study, for example,
1) the metallicity gradients and distributions of the different stellar populations of the Milky Way,
2) the structure of the Galactic disk(s) as a function of metallicity and spectral type,
3) the binary fraction of field stars as a function of stellar color and metallicity (Yuan et al. 2015d),
4) the selection effects of spectroscopic surveys such as the LAMOST Galactic surveys,
and 5) providing detailed constraints on Galactic models such as the Besancon model (Robin et al. 2003). 

The light-motion curve catalog of Bramich et al. (2008)
contains proper motions for about 1 million stars down to a limiting magnitude of $r \sim$ 21.5\,mag, 
thus includes the majority sample stars of the current work.
With proper motions from Bramich et al. (2008) and radial velocity measurements from LAMOST and other facilities, 
the current sample provides a valuable dataset to 
study the kinematics (e.g., Smith et al. 2009a) and 
sub-structures of the Galactic disk and halo (e.g., Smith et al. 2009b), 
and to constrain the local dark matter density (e.g., Zhang et al. 2013).

The current sample probes mainly the inner disk and halo of the Galaxy, 
thus is complementary to the LAMOST Spectroscopic Survey of the Galactic Anti-center (Liu et al. 2014; Yuan et al. 2015a)
that targets the Galactic outer disk. The sample is accessible from http://162.105.156.249/site/Stripe82/ along
with a descriptive readme.txt file.

\vspace{7mm} \noindent {\bf Acknowledgments}{
{We would like to thank the referee for his/her useful comments.}
This work is supported by National Key Basic Research Program of China 2014CB845700,
NSFC grant 11443006, and China Postdoctoral Science special Foundation 2014T70011.
This work has made use of data products from the Sloan Digital Sky Survey.

Funding for SDSS-III has been provided by the Alfred P. Sloan Foundation, the
Participating Institutions, the National Science Foundation, and the U.S.
Department of Energy Office of Science. The SDSS-III web site is
http://www.sdss3.org/.

SDSS-III is managed by the Astrophysical Research Consortium for the
Participating Institutions of the SDSS-III Collaboration including the
University of Arizona, the Brazilian Participation Group, Brookhaven
National Laboratory, Carnegie Mellon University, University of Florida, the
French Participation Group, the German Participation Group, Harvard University, the
Instituto de Astrofisica de Canarias, the Michigan State/Notre Dame/JINA
Participation Group, Johns Hopkins University, Lawrence Berkeley National
Laboratory, Max Planck Institute for Astrophysics, Max Planck Institute for
Extraterrestrial Physics, New Mexico State University, New York University,
Ohio State University, Pennsylvania State University, University of Portsmouth,
Princeton University, the Spanish Participation Group, University of Tokyo,
University of Utah, Vanderbilt University, University of Virginia, University
of Washington, and Yale University.
}

\end{document}